\documentclass[prl,twocolumn,showpacs,amsmath,amssymb]{revtex4}

\input epsf
\usepackage{graphicx}
\usepackage{dcolumn}
\usepackage{bm}

\begin{document}

\preprint{}

\title{Local Stress Relaxation and Shear-banding in a Dry Foam under Shear}

\author{Alexandre Kabla}
\author{Georges Debr\'egeas}
\email{georges.debregeas@college-de-france.fr}
\affiliation{LFO - Coll\`ege de France, CNRS UMR $7125$, Paris, France}
\date{\today}

\begin{abstract}

We have developed a realistic simulation of 2D dry foams under
quasi-static shear. After a short transient, a shear-banding
instability is observed. These results are compared with
measurements obtained on real 2D (confined) foams. The numerical model 
allows us to exhibit the mechanical response of the material to a single
plastication event. From the analysis of this elastic propagator,
we propose a scenario for the onset and stability of the flow
localization process in foams, which should remain valid for most
athermal amorphous systems under creep flow.

\end{abstract}


\maketitle

Amorphous glassy materials are ubiquitous in industry and nature:
they include silica-based glass-formers and polymer melts below
$T_g$, dense colloidal suspensions and emulsions, foams and dense
granular systems. Unlike crystalline solids, plasticity in such
systems originates from discrete {\it local} relaxation events
\cite{Bulatov,Falk}, involving a small number of particles (atoms, grains,
bubbles, etc...). Spatial and time correlations in the occurrence
of these plastic events are generally important, leading to
avalanche-like dynamics \cite{Obukov,Tewari1999,Jiang1999} and
spatially inhomogeneous flows \cite{Lequeux,Mueth2001,Langer2001}. 
Glassy rheology thus remains one of the most active and
challenging domains of statistical physics. 

Amongst the large number of theoretical and numerical models
recently proposed, foam has emerged as a 
strongly inspiring model system \cite{Sollich1997, Durian1997, Ono2002}.
First, because thermal energy is
strictly irrelevant on bubble scale, creep flow experiments can be
run (by imposing an infinitely low deformation rate) in which time
dependent effects are absent. Second, the bubble mechanics is
simple and yields a wide linear elastic regime. Finally,
plasticity in foams is associated with well identified processes.
In spite of this apparent simplicity, many features of foams
flow remain to date unexplained \cite{WeaireBook}. 
Thus, shear-banding flows have been recently exhibited 
in a 2D Couette experiment
\cite{Debregeas2001}. In this study, a monolayer of bubbles
squeezed between two glass plates was slowly sheared between two
concentric discs. Much of the rearrangements were found to occur
in a thin region (a few bubbles in width) along the edge of the
inner disc.
In the present letter, we directly address this question by
developing a numerical model adapted to the quasi-static shearing
of 2D dry foams. The observed flow features are compared with experimental data obtained with the same set-up as in \cite{Debregeas2001}. This
model allows us to investigate the micro-scale mechanics of the foam
leading to strain localization. 

\begin{figure}
\centerline{ \epsfxsize=8.5truecm \epsfbox{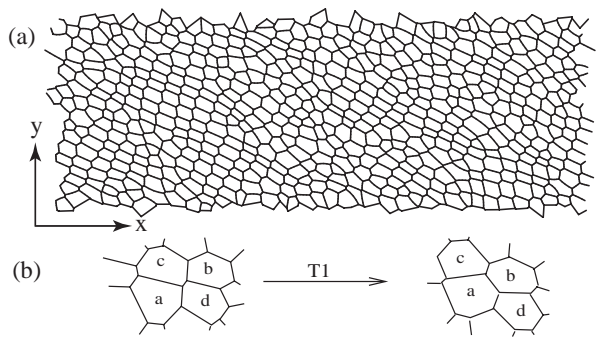}}
\caption{\label{figure1} (a) Snapshot of a simulated foam with $16
\times 48$ bubbles. The polydispersity is 6\%. The foam has
periodic boundary conditions along the $x$ direction. Films laying
at upper and lower edges are fixed; shearing is obtained by moving
the lower edge along the $x$ direction.
(b) Example of a topological change ($T1$ process) occurring 
inside the foam upon shearing.}
\end{figure}

We use Voronoi tessalation to build polydisperse structures of $W
\times L=16 \times 48$ cells separated by straight segments.
These are later referred as bubbles and
films respectively, by analogy with real foams, the intercept
between films being called a vertex. These structures have
periodic boundary conditions along the $x$ direction, and films
laying at the upper and lower edges are fixed to allow subsequent
plane parallel shearing (Fig. \ref{figure1}(a)).  
To obtain a mechanically equilibrated structure, 
the total film length is minimized at fixed topology and with a constant 
volume constraint on each bubble, as
expected for static dry foams \cite{Weaire1983}.
Our algorithm is based on Surface Evolver \cite{SE}, a
software widely used in foams structural studies \cite{Kraynik}.
The main difficulty of this minimization procedure  
comes from the existence of very soft modes associated
with large-scale shear deformations \cite{Herdtle}.
A special care is thus put in equilibrating these modes. 
The overall procedure is then validated by
imposing various strain fields to the initial foam, and
checking that the resulting equilibrated structure
remains unchanged.

Once the foam has been mechanically equilibrated, plasticity 
is introduced by allowing $T1$ rearrangements -
the elementary topological changes in 2D foams (see Fig.
\ref{figure1}(b)).  In a real dry foam, vertices have 
a finite size which depends on the liquid fraction. 
When a film becomes smaller than this length,
the two vertices attract and a $T1$ event is triggered. 
We mimic this criterion by exchanging bubbles neighbors when one 
of the film length falls below a fixed value $l_v$, corresponding 
to a liquid fraction $\phi = 1\%$. The $T1$ events are triggered
 one at a time and followed by a complete mechanical equilibration. 
This two-step procedure is iterated until all films are stable with 
regards to the $T1$ criterion. It should be noted that this procedure
might not precisely reflect the physical process taking place 
during an avalanche of $T1$ events. Indeed, in a real 
foam plasticity and mechanical equilibration take place simultaneously.
Our procedure implicitly assumes the latter to be much faster
than the $T1$ event.
Finally, the foam is quasi-statically sheared by iteratively
moving the lower edge over a short distance then equilibrating the structure and allowing plastic events.

\begin{figure}[htb]
\centerline{ \epsfxsize=8.5truecm
\epsfbox{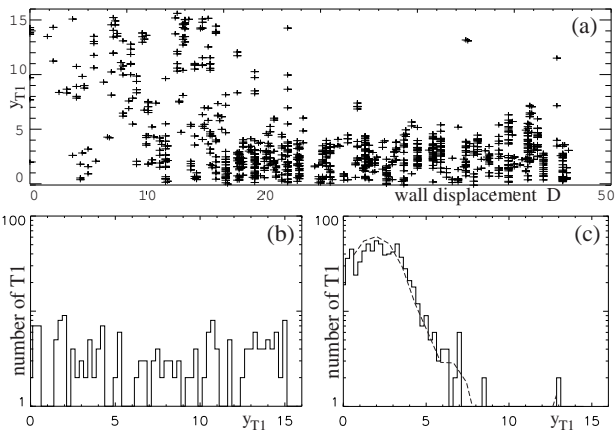}}
\caption{\label{figure2} Localization process in a simulated foam.
(a) The $y-$position of the $T1$ events as a function of the wall
displacement $D$ expressed in bubble diameter. (b) Distribution of
the $y-$positions of the $T1$ events for $D<15$ (transient regime)
and (c) $D>20$ (localized regime).  In the latter, the dotted line
shows the gradient $\frac{\partial \overline v}{\partial y}(y)$ of
the associated plastic flow profile $\overline v(y)$.}
\end{figure}

Figure \ref{figure2} conveys the main result of the present study:
it displays the $y_{T1}$ positions of the rearrangements as a
function of the imposed wall displacement $D$. After a short
transient (for $D \sim W$ {\it i.e.} an imposed strain $\sim 1$),
the rearrangements permanently gather within a thin shear-band in
the vicinity of the lower wall. This strain instability is
observed for all the simulations, with a flow localization taking
place on either wall depending on the initial foam structure. In the
following we mainly focus on measurements performed in
the steady-state localized regime.

From the sequence of equilibrated structures, we measure the
trajectories of the bubbles centers to extract the flow field at
each time step. Furthermore, we can compute the internal shear
stress on any sub-volume $w$ using the following relation (where
the summation is performed over all segments $\vec{l}$ inside $w$)
\cite{WeaireBook}:

\begin{equation}
\sigma_{xy}(w) = \frac{1}{w} \cdot \sum_{\vec{l}\in w}
\frac{l_{x} \cdot l_{y}}{l}
 \label{stress}
\end{equation}
\vspace{-0.3cm}

\begin{figure}[htb]
\centerline{ \epsfxsize=8.5truecm \epsfbox{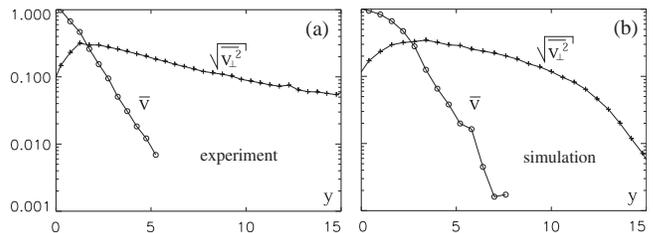}}
\caption{\label{figure3} comparison of time averaged measurements
obtained from (a) experiments on 2D foams in a Couette cell (b)
simulated foams. $\overline v$ is the tangential velocity rescaled
by the wall velocity $v_0$. $\sqrt{\overline{{v_\perp}^2}}$ is the
normal mean square displacement, associated with a time-lapse
$\tau=0.25 \; d/v_0$, where $d$ is the average bubble diameter.
The rapid drop of $\sqrt{\overline{{v_\perp}^2}}$ in (b) far from
the moving wall is due to the relatively small width of the
simulated foam ($W=16$), and hence the presence of the other
confining wall.}
\end{figure}

We have compared time averaged measurements from the simulation
with experimental data we obtained using the same liquid fraction
($\phi = 1\%$).
Figure \ref{figure3} shows the tangential velocity
profiles and the normal velocity fluctuations  for the experiment
and the simulation. For both quantities, we observe similar decays
with the distance from the wall. Other flow features, such as the
stress fluctuations profiles (presented below),
show a similarly good agreement. This adequacy proves the validity
of the present simulation. Conversely, it demonstrates that the
shear-banding observed in \cite{Debregeas2001} is not due to the
Couette geometry, in which the mean stress
decreases with the distance from the inner disc.\\

Beyond these time averaged measurements, the simulation allows one
to study the evolution of the foam on short time scales. The
dynamics can be separated into two elementary processes,
associated with different simulation time steps: (i) charge
periods over which the position of the wall is incremented without
plastication. The resulting deformation is linear and the shear stress
tensor uniformly increases. This allows us to extract a shear
elastic modulus $\mu$. This modulus is found to weakly depend
on the total applied strain and is considered as a constant in the following. 
(ii) plastic yielding, during which the
stress is relaxed through discrete $T1$ events. To analyze in
detail the latter, we focus on the displacement and shear stress
fluctuation fields produced by a single rearrangement. The spatial
resolution is enhanced to below one bubble diameter by averaging
these results over $~100$ individual $T1$ events located at the
same $y_{T1}$ coordinate.

\begin{figure}
\centerline{ \epsfxsize=8.5truecm \epsfbox{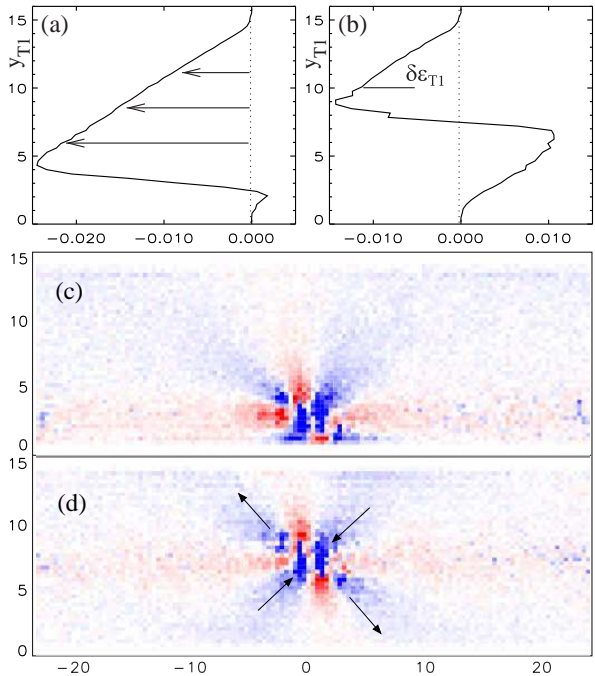}}
\caption{\label{figure4} Displacement and variational shear stress
field associated with a single $T1$ event. Each of these results
have been obtained by averaging over $\sim 100$ different $T1$
events located at the same distance $y_{T1}$ from the shearing
wall. (a) and (b): line averaged displacement profiles (expressed
in bubble diameter $d$), for a $T1$ event located at $y_{T1}=3$
and $8$ respectively. $\delta \epsilon_{T1}$ represents the mean
strain released by the $T1$ event. (c) and (d): Corresponding
shear stress variation fields. Red color indicates an increase of
the stress (relative to the imposed shear stress), blue color
indicates a stress relaxation. The arrows show approximately the
motion of the rearranging bubbles during the $T1$ event.}
\end{figure}

Figure \ref{figure4} (a) and (b) display the average displacement
profiles associated with $T1$ events located at two different
distances $y_{T1}$ from the lower wall. Both profiles exhibit a
strong discontinuity at the rearranging line whereas 
the rest of the material is uniformly deformed with a strain
amplitude $\delta \epsilon_{T1}$. Regardless of the position
$y_{T1}$, $\delta \epsilon_{T1}=1.07 \; d^2/(WL)$ (with a $30\%$
statistical dispersion over different $T1$ events), where $d$ is
the mean bubble diameter and $WL$ is the foam area.

This elementary strain $\delta \epsilon_{T_1}$ can be interpreted
from two different viewpoints. On one hand, it represents a {\it
plastic} strain amplitude: each $T1$ event increments the plastic
flow gradient at $y=y_{T1}$ by $-\delta \epsilon_{T1}$ in average.
This yields the following kinematic relation between the plastic
flow profile $\overline v(y)$ and the spatial distribution of $T1$
events:

\begin{equation}
 \frac{\partial \overline v}{\partial y}(y) = -W \; \omega(y) \;
 \delta \epsilon_{T_1}
 \label{cinemat}
\end{equation}
\vspace{-0.3cm}

where $\omega(y)dy$ is the frequency of $T1$ events occurring
between $y$ and $y+dy$. This relation can be directly exhibited by
over-plotting the plastic velocity gradient on the $T1$ spatial
distributions (see Fig. \ref{figure2}(c)). On the other hand,
$\delta \epsilon_{T_1}$ is a uniform elastic strain relaxation.
The associated stress can be independently evaluated using Eq.
(\ref{stress}) yielding a line-averaged uniform stress release
$\delta \sigma_{T1}=\mu \delta \epsilon_{T1}$. By taking into
account both the elastic charge and $T1$ relaxation, we derive an
equation of evolution of the line-averaged shear stress
$\overline{\sigma}(y)$, valid for any line $y$ between $0$ and $W$:

\begin{equation}
\label{stress} \mu \dot{\gamma}-\mu  \delta \epsilon_{T1}
\int_{0}^W{\omega(y') dy' = \frac{\partial  {\overline{\sigma}
(y)}}{\partial t} = 0}
\end{equation}

The first term of the left-hand side of the equation corresponds
to the advective charge induced by the imposed shear at strain
rate $\dot{\gamma}$. The second term comes from the cumulative
relaxation of stress associated with the $T1$ processes. The
integral form of this equation is a direct consequence of the long
range mechanical relaxation associated with each $T1$ process. As
a result, this line-averaged mechanical analysis can not allow one
to predict a flow profile, and is in fact strictly equivalent to
Eq. (\ref{cinemat}) from which it can be deduced by simple
integration. In other words, any velocity profile which obeys the
kinematic boundary conditions is mechanically admissible.
\\

The understanding of the shear-banding instability finally comes
down to the following question: All lines bearing in average the
same stress, why are $T1$ events unevenly distributed amongst
them? To capture this process, we need to go beyond the line-averaged
analysis and examine the spatial structure of stress release associated
with individual T1 events. This is shown in Fig. \ref{figure4} (c) and (d), 
for two different $y_{T1}$ locations. As it appears clearly, 
the stress release is very inhomogeneous and anisotropic, 
owing to the systematic displacement pattern of the rearranging bubbles 
imposed by the
shearing \cite{CASTEM, Langer2001}. In particular, lines in the vicinity of
the rearranging site experience large stress modifications.
Although the global effect is a release of the main shear stress,
some regions (which appear in red) get over-charged. By contrast,
remote lines are homogeneously relaxed. 

\begin{figure}[htb]
\centerline{ \epsfxsize=8.5truecm \epsfbox{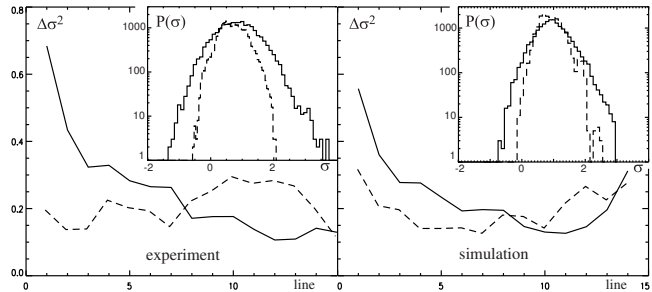}}
\caption{\label{figure5} 
Profile of shear stress variance $\langle (\sigma(x,y)-\overline{\sigma})^2 \rangle$ 
in real (left) and simulated (right) foams. Solid lines correspond to foams in the
fully localized regime. Dashed lines correspond to freshly prepared samples (no shearing). 
All data have been rescaled with the long time limit shear stress. Inset : shear stress
probability distributions at lines $y=1$ (solid) and $y=10$ (dashed) respectively, in
the localized regime.}
\end{figure}

From this measurement, one may expect that T1 events do not only relax the
global stress but also cumulatively modify the statistical 
properties of the frozen stress field.
To investigate this effect, we
measure the shear stress distributions $P(\sigma(x,y))$ at different distances $y$
from the shearing wall. We extract from these distributions the local variances 
$\Delta \sigma^2(y)=\langle (\sigma(x,y)-\overline{\sigma})^2 \rangle$. We then 
compare these profiles obtained from foam structures
before shearing and after full localization.
As shown in Fig. \ref{figure5}, the shearing induces an inhomogeneous modification
of these profiles:
a large increase of $\Delta \sigma^2$ occurs in the shear band 
region where many T1 events have occurred, in both the experiment and the simulation.
By contrast, the stress distributions in lines
away from the shear band display no modification,
or even a small decrease of their variance. The latter is due to T1's 
occurring during the transient period of charge which do not have a 
systematic orientation. We therefore postpone the discussion 
of this effect.

This result shows that the strain history of the foam is permanently 
imprinted in its frozen stress field, and that such modification
can be directly probed through measurements of  $\Delta \sigma^2$.  
This parameter has a further important physical meaning: 
large values of $\Delta \sigma^2$ indicate that a large fraction of bubbles 
are highly deformed and therefore more likely to rearrange upon increasing the global stress.
This parameter thus provides a local measurement of the foam ``fragility''.

Based on these observations, a simple scenario for
strain localization in quasi-static shearing can be proposed.
Starting with a homogeneous structure, shear-banding develops through
a self-amplification process: $T1$ events locally weaken the foam
structure by increasing its frozen stress disorder. This in turn enhances
the probability for subsequent rearrangements to take place 
in neighboring lines. This mechanism spontaneously leads to the
formation of a single shear-band in the material. 
Within this scheme, we can  also qualitatively understand 
why shear-bands preferentially
develop along the boundaries, even in plane parallel shearing 
geometry where the average shear stress is uniform. 
Indeed, the presence of a rigid boundary with a no-slip condition imposes
an extra mechanical constraint to the foam in the vicinity of the walls. This
 tends to locally enlarge the local stress distributions.
\\

The experimental and numerical systems studied here provides an ideal 
model to study plasticity in disordered media. It allowed us to access 
detailed mechanical features, from the stress signature associated 
with a single plastic event, to the statistical modifications of the 
frozen stress field associated with a fully developed
shear flow. We have used these results to propose a simple scenario 
for shear localization based on a strain weakening process. 
Most results obtained with this model system should remain
valid to any material provided the existence of
(i) frozen disorder (no thermal relaxation),
(ii) elastic behavior at low deformation, (iii) local discrete plastication processes.
It could therefore serve as a useful test 
to more elaborated models of plasticity 
that involve local stress relaxations \cite{Bulatov,Langer2001, Derec2001}
but do not necessarily address foam rheology. 

\begin{acknowledgments}
We wish to thank J. Scheibert, C. Caroli, O. Pouliquen and 
J.-M. di Meglio for stimulating discussions.
We are grateful to C. Fond for introducing us to 
finite elements calculation.
\end{acknowledgments}

\end{document}